\documentstyle[12pt,epsf,epsfig]{article}
\def\be{\begin{equation}}
\def\ee{\end{equation}}
\def\ba{\begin{eqnarray}}
\def\ea{\end{eqnarray}}
\def\ga{\mathrel{\raise.3ex\hbox{$>$\kern-.75em\lower1ex\hbox{$\sim$}}}}
\def\la{\mathrel{\raise.3ex\hbox{$<$\kern-.75em\lower1ex\hbox{$\sim$}}}}
\newcommand{\sect}[1]{\section{#1}\setcounter{equation}{0}}

\newcommand{\bi}[1]{\bibitem{#1}}

\begin{document}
\baselineskip=16pt
\begin{titlepage} 
\rightline{UMN--TH--1901/00}
\rightline{TPI--MINN--00/23}
\rightline{hep-ph/0005146}
\rightline{May 2000}  
\begin{center}

\vspace{0.5cm}

\large {\bf Solving the Hierarchy Problem in
Two-Brane Cosmological Models
}
\vspace*{5mm}
\normalsize

{\bf  Panagiota Kanti, Keith A. Olive} and {\bf Maxim Pospelov}

\smallskip 
\medskip 
 
{\it Theoretical Physics Institute, School of Physics and
Astronomy,\\  University of Minnesota, Minneapolis, MN 55455, USA} 

\smallskip 
\end{center} 
\vskip0.6in 
 
\centerline{\large\bf Abstract}

We analyze cosmological solutions in the class of 
two-brane models with arbitrary tensions which contain matter
with general equations of state.  We show that the mass
hierarchy between the two branes is determined by the ratio of
the lapse functions evaluated on the branes. This ratio can be
sufficiently small  without fine-tuning the brane separation, 
once the transverse  dimension is stabilized. For suitably large
interbrane separations, both brane tensions are positive. We also find
that the cosmological evolution obeys the standard four-dimensional 
Friedman equation up to small corrections.

\vspace*{2mm} 

\end{titlepage}

\sect{Introduction}

Physical consequences of extra space-like dimensions have been
the subject of intensive theoretical investigations 
over the past two years. In general,
the presence of new dimensions can change our interpretation of the 
fundamental parameters in the theory. Four-dimensional constants, such 
as Plank mass, electroweak scale, coupling constants and so on, 
are, in fact, functions of the volume of the extra dimensions. This
dependence  is particularly interesting in theories where gravity and the
rest of  the Standard Model fields ``live'' in different numbers of
dimensions.  This can be realized in the brane-world scenario, in which
Standard Model particles are confined to a 3+1 space-time dimensional
manifold,  embedded into a higher-dimensional space. A very well known
stringy prototype
of these constructions \cite{w}, uses a relatively large size of the 
transverse 11-th dimension in supergravity to reconcile effective Plank 
scale and GUT/string scale. In such a construction the 4D-Planck
scale, $M_P$, is derived from the fundamental scale, $M_{11}$, using
$M_P^2 = M_{11}^3 R_{11}$ having assumed that the 6-space volume,
$V_6 \sim M_{11}^{-6}$. For $R_{11} \gg M_{GUT}$, we can obtain, 
$M_{GUT} \sim M_{11} \ll M_P$, thus relaxing the hierarchy problem of
the GUT and string scales. 

Interest in such models, driven by particle phenomenology and cosmology,
was further amplified when the brane world idea was
used to  lower the fundamental gravitational scale, bringing it close to
the  weak scale \cite{ADD} in an attempt to resolve the hierarchy problem
between the weak and Planck scales. Indeed, as it was shown in
\cite{ADD}, due to a possibly very large volume of extra dimensions felt
only by gravity, the fundamental scale could be much lower 
than the one inferred from measuring the Newtonian force in four
dimensions.
This opens up a number of interesting phenomenological consequences,
including the possibility of observing higher-dimensional gravitational
interactions in collider experiments.
An alternative to solve the weak scale hierarchy problem was proposed by
Randall and Sundrum \cite{RS1}. Rather than requiring a large extra
dimension, one simply constructs an exponentially large ratio of the two
length scales  on two different branes, which results from an
``exponentially decaying'' scale factor (or warp factor) in AdS geometry.
For other attempts at solving the hierarchy problem with extra dimensions
see \cite{others}. 

The study of the cosmological evolution in this type of model has
undergone several interesting turns. It was shown in
\cite{LOW,LK,BDL} that  the Hubble parameter on the brane is proportional
to the brane tension,  rather than to the square root of it, as one would
expect from pure low-energy four-dimensional reasoning. Two-brane models
of this type have additional problems since the matter densities $\rho_1$,
$\rho_2$ and pressures
$p_1$ and $p_2$ need to be correlated in order to achieve 
a consistent solution (in fact one would derive that $\rho_2 \simeq
-\rho_1$ which lacks a consistent interpretation). This also leads to the
correlation  between the states of matter on   the two different branes.
These problems with cosmology are indicative that some crucial ingredients
are missing in the brane models  which would allow a smooth
transition to ``normal'' four-dimensional physics. It was 
conjectured in \cite{kkop1} and further proven in \cite{randall3,kkop2} 
that the missing element in brane models is the stabilization of 
the extra dimension.

It was shown in Refs. \cite{kkop1,kkop2} that single-brane models
with 
a compactified extra dimension lead to standard cosmology, irrespective of
the nature of the precise mechanism responsible for the stabilization of the 
transverse direction. In technical terms, stabilization provides 
a bulk value for  the transverse pressure-like component of 
the stress-energy tensor, proportional to the 
trace of the stress-energy tensor of the brane. By including the bulk
component of energy-momentum tensor ($T_{55}$), one not only recovers
the standard form of the Friedmann equation, but also relieves the
necessity of the (anti) correlation of the energy density on the brane
allowing for the construction of realistic two- and one-brane models.
In \cite{kkop2}, it was shown that $T_{55}$ is automatically generated as
a back-reaction to matter on the brane and induces a shift (proportional
to $T^\mu_\mu$) in the minimum of the dilaton potential. However, in such
models with matter, the warp factor is insignificant (it is proportional
to
$\rho/M_P^4$) and we lose the potential solution to the hierarchy problem. 

In our subsequent paper 
\cite{kop1}, we generalized this construction and 
found a static solution for the case 
of two positive tension branes, assuming the same stabilization
mechanism.
An interesting feature of this construction, noted in \cite{kop1}, is the 
possibility of the exponentially large ratio between the 
two scales on two different branes, which allows to address the gauge 
hierarchy problem a l\`a Randall and Sundrum (RS)\cite{RS1}, and at the same
time 
avoid the presence of the negative tension brane. 
Two positive tension branes are allowed in this context because the
solution to the warp factor is $cosh$-like rather than a pure
exponential. Even though the 
negative tension branes are plausible in string theory, their existence in
the 
real physical world is questionable. Indeed, these branes may turn out to
be
unstable, once the supersymmetry is broken, and to the best of our 
knowledge this question has not been properly studied so far. In
contrast, positive tension branes are ubiquitous, and occur 
not only in string theory, but also in an ordinary field theories 
where they can simply be interpreted as domain walls, domain wall 
junctions, etc.

We believe that this idea \cite{kop1} deserves further 
consideration, and in this paper
we take it one step further, introducing matter density and pressure 
on the two branes in order to obtain consistent cosmological solutions. In
accordance  with what we learned in the case of the single-brane models,
this expansion 
is exactly of the Friedmann form, provided that the transverse dimension is 
stabilized. At the same time, there is no correlation between $\rho_1$,
$\rho_2$, $p_1$ and $p_2$, and no restriction on the equations of state. 
We obtain general solutions for arbitrary brane tensions and we further
demonstrate that it is possible to recover a solution to the hierarchy
problem for negative or positive values of the observable brane tension.
As a result, the general system of two branes discussed in this paper
gives rise to cosmological solutions that include, as special cases,
both the positive-negative brane combination of the RS model \cite{RS1}
and the more natural case of the two positive tension branes \cite{kop1}.
We find that
the system is characterized by the presence of three parameters:
the ratio of the self-energy over the bulk cosmological constant of
each brane and the inter-brane distance. Demanding the
cancellation of the effective cosmological constant, we place one
constraint on a combination of these parameters while the resolution
of the hierarchy problem fixes one more parameter. We keep the
inter-brane distance, $L$, as the free parameter and we demonstrate
that suitably small values of the warp factor, which we show to be
directly related to the ratio $M_P/M_W$, is achieved for any value of
$L$. In other words, no other fine tuning is necessary to obtain the
necessary warp factor which we show to be directly related to the ratio
of the lapse functions on the two branes.

As indicated above, we find a normal Friedmann expansion on the visible
brane and show that in fact both the visible and hidden branes
expand with the {\em same} expansion rate. Upon rescaling the matter
density on the visible brane, we further show that the deduced Plank
scale is the same on both branes. Although the scale factors and lapse
functions change in time leading to a time dependent ratio of weak 
scale and four-dimensional Plank constant, we  show
that the residual change in
$M_W/M_{P}$ is minute, suppressed by the 
ratios of $\rho_i/ M_{P}^4$, and thus representing no danger to this
model.
 
This paper is organized as follows. In section 2, we introduce 
the ansatz for the metric, stress-energy tensor and present the
five-dimensional Einstein's equations. In the next section, we derive
the solution for the spatial scale factor, in the bulk, consistent
with the brane boundary conditions, and show that the
cosmological expansion rate on the two branes is of the form of the
standard four-dimensional Friedmann equation, once the inter-brane
distance is stabilized. In section 4, we consider the solution to the
hierarchy problem with branes with arbitrary tensions, and define the
necessary conditions for this to happen.  We demonstrate that the
hierarchy problem can be resolved for arbitrary values of the inter-brane
separation, by appropriately choosing the value of the observable brane
tension, with the RS choice of parameters arising only as a special case.
We draw our conclusions in section 5.

\sect{The Field Equations}

The starting point of our analysis will be the assumption of the
existence of an extra dimension, denoted by the coordinate $y$, in
addition to the usual four coordinates, $\{t,x^i\}$, of the 4-dimensional
spacetime. We consider the following ansatz for the line-element of
the 5-dimensional manifold
\be
ds^2=-n^2(t,y)\,dt^2 + a^2(t,y)\,\delta_{ij} dx^i dx^j + b^2(t,y)\,dy^2\,,
\label{metric}
\ee
where $a(t,y)$ and $b(t,y)$ are the scale factors of the 4-dimensional
spacetime (which we assume is spatially flat, i.e., $^3k = 0$) and the
extra dimension, respectively, and
$n(t,y)$ the lapse function that defines the time variable. Motivated by
M-theory~\cite{w}, we assume that the extra dimension has the topology
of an orbifold, i.e. is a circle where the discrete $Z_2$ symmetry,
$y \leftrightarrow -y$, has been imposed.  
Two 3-branes of zero thickness located at the
orbifold's fixed points, $y=0$ and $y=L$, set the size of the extra
dimension to be equal to $2bL$. The inter-brane distance is assumed 
to remain fixed due to a stabilization mechanism which ensures that
$\dot b=b'=0$. The coordinate $y$ is scaled so that we may set, for
simplicity, $b=1$. Note that, throughout the paper, dots and primes denote
differentiation with respect to time and $y$, respectively. For other
approaches to brane-world cosmologies, see \cite{morerefs}.

We now turn to the 5-dimensional action that describes the
coupling of the matter content of the universe with gravity. We assume
that all the usual matter fields are localized on the two 3-branes, at
$y=0$ and $y=L$, that play the role of a hidden and observable universe,
respectively. The two branes are characterized by self-energies $\Lambda_1$
and $\Lambda_2$, while a non-vanishing cosmological constant
$\Lambda_B$ exists in the bulk. Under the above assumption, the action
takes the form
\be
S=-\int d^{4}x\,dy\,\sqrt{-\hat{g}}\,\biggl\{\frac{
M_{5}^{3}}{16\pi }\,\hat{R}+\Lambda _{B} +\Lambda _{1}\,\delta
(y)+\Lambda _{2}\,\delta (y-L)+  \hat{\cal L}_o\biggr\}\,.
\label{action0}
\ee
In the above, $\hat{\cal L}_o$ represents all possible contributions to
the action which are not strictly gravitational, $M_{5}$ is the
fundamental 5-dimensional Planck mass and the hat denotes 5-dimensional
quantities.  

Einstein's equations,
with the above spacetime background (\ref{metric}) and the
assumption that $\dot b=b'=0$, take the simplified form (for the full
version of these equations see e.g. Refs. \cite{LOW,BDL})
\ba
&~& \hspace*{-0.5cm} \hat{G}_{00} = 
3\,\Biggl\{\biggl(\frac{\dot{a}}{a}\biggl)^2
-\, n^2\,\Biggl[\frac{a''}{a} + 
\biggl(\frac{a'}{a}\biggr)^2 \Biggr]\Biggr\}
= \hat{\kappa}^2 \, \hat{T}_{00}\,,\label{00}\\[4mm] 
&~&\hspace*{-0.5cm} \hat{G}_{ii} = a^2\,\Biggl\{\frac{a'}{a}\,
\Biggl(\frac{a'}{a} + 2\frac{n'}{n}\Biggr) +2 \frac{a''}{a}
+\frac{n''}{n}\Biggr\}
+ \frac{a^2}{n^2}\,\Biggl\{\frac{\dot{a}}{a}\,\Biggl(-\frac{\dot{a}}{a} +
2\frac{\dot{n}}{n}\Biggr) -2\frac{\ddot{a}}{a}\Biggr\}= 
\hat{\kappa}^2\,\hat{T}_{ii}\,,
\label{ii}\\[4mm]
&~& \hspace*{-0.5cm} \hat{G}_{05} = 3\,\Biggl(\frac{n'}{n}
\frac{\dot{a}}{a} -\frac{\dot{a}'}{a}\Biggr)=0\,,
\label{05}\\[4mm]
&~& \hspace*{-0.5cm} \hat{G}_{55} = 
3\,\Biggl\{\frac{a'}{a}\,\Biggl(\frac{a'}{a} +
\frac{n'}{n}\Biggr) -\frac{1}{n^2}\,\Biggl[\frac{\dot{a}}{a}\,
\Biggl(\frac{\dot{a}}{a}-\frac{\dot{n}}{n}\Biggr) +
\frac{\ddot{a}}{a}\Biggr]\Biggr\} = \hat{\kappa}^2\,\hat{T}_{55}\,, 
\label{55}
\ea
where $\hat{\kappa}^2=8\pi \hat{G}=8\pi/M_5^3$. In the above,
$\hat T_{MN}$ is the total 5-dimensional energy-momentum tensor which
can be decomposed in terms of the bulk and brane contributions as
follows
\be
\hat T_{MN}= \hat T_{MN}^{(B)} + \hat T_{MN}^{ (1)} +
\hat T_{MN}^{ (2)}\,,
\ee
where 
\ba
&~& \hat{T}^M_{(B)\,N}={\rm diag} (- \Lambda_B, - \Lambda_B,
- \Lambda_B, - \Lambda_B,\hat{T}^5_{(B)\,5})\,, \nonumber \\[3mm]
&~&  \hat{T}^M_{(1)\,N}={\rm diag} \biggl[\frac{\delta(y)}{b}\,
\Bigl(- {\rho_1 -\Lambda_1}, {p_1 - \Lambda_1}, {p_1 - \Lambda_1},
{p_1 - \Lambda_1}\Bigr), 0\biggr]\,,\\[3mm]
&~&  \hat{T}^M_{(2)\,N}={\rm diag} \biggl[\frac{\delta(y-L)}{b}\,
\Bigl(- {\rho_2 - \Lambda_2}, {p_2 - \Lambda_2}, {p_2 - \Lambda_2},
{p_2 - \Lambda_2}\Bigr), 0\biggl]\,. \nonumber
\ea
Note that we allow for a non-vanishing value of the (55)-component
of the energy-momen\-tum tensor in the bulk that includes, in addition
to the contribution due to the bulk cosmological constant, a part
proportional to the trace of the energy-momentum tensor on the brane.
As it has been shown~\cite{kkop1,randall3,kkop2}, this component
results from a 5-dimensional stabilization mechanism which is
responsible for keeping the inter-brane distance, and thus the
size of the extra dimension, fixed. The presence of a stabilizing
potential for the radion field in the framework of the 5-dimensional,
fundamental theory leads to a $(t,y)$-dependent value for
$\hat{T}_{(B)\,5}^{5}$ distinctly different from $-\Lambda _{B}$.
In particular, it can be shown that $\hat{T}_{(B)\,5}^{5} + \Lambda_B$ is
(to leading order in $\rho$) proportional to the trace $T^\mu_\mu$ on the
brane. For later use, we display here the relations that follow from the
conservation of the energy-momentum tensor,
$D_M \hat{T}^M_{\,\,\,\,\,\,N}=0$, on the branes
\be
\frac{d \rho_1}{dt} + 3(\rho_1+ p_1)\,\frac{\dot{a}_0}{a_0} =0\,,
\qquad 
\frac{d \rho_2}{dt} + 3(\rho_2+p_2)\,\frac{\dot{a}_L}{a_L} =0\,,
\label{zeroth} 
\ee
and in the bulk
\be
\Bigl(\hat{T}^5_{(B)\,5}\Bigr)^{'} + \hat{T}^5_{(B)\,5}
\Biggl(\frac{n'}{n} + 3\frac{a'}{a}\Biggr) +
\Lambda_B\,\Biggl(\frac{n'}{n} + 3\frac{a'}{a}\Biggr)= 0\,.
\label{fifth}
\ee

Here, we closely follow the analysis of Ref. \cite{kkop2} which we
extend to the case of two branes with arbitrary self-energies
$\Lambda_i$'s. Our subsequent analysis will be greatly simplified
by the fact that the (05)-component of Einstein's equations (\ref{05})
can be easily integrated to give the result
\be
n(t,y)=\lambda(t)\,\dot{a}(t,y)\,.
\label{soln}
\ee
Using the normalization $n(t,y=0)=1$, the arbitrary function of
time $\lambda(t)$ turns out to be $\lambda(t)=1/\dot a_0$ in terms of
which, as we will shortly see, the Hubble parameter on both branes can be
easily expressed.


\section{Cosmological Evolution}

In this section, we focus on the general solution
for the scale factor $a(t,y)$ which is consistent with the aforementioned
energy distribution in the 5-dimensional universe and the boundary
conditions that the two, infinitely thin, brane-universes introduce in
the theory. We will also determine the form of the generalized
Friedmann equations that the boundary quantities $a_0$ and $a_L$ satisfy
and we will investigate the conditions necessary for the restoration of
the usual Friedmann equation on the two branes, through the vanishing of
the effective cosmological constant, as well as for the successful
stabilization of the extra dimension. 

We note that by using the relation (\ref{soln}), the $(00)$-component
of Einstein's equations (\ref{00}), in the bulk, reduces to an ordinary
second-order differential equation for $a(t,y)$ with respect to $y$. 
As a result, we can easily find the general solution for the scale factor,
in the bulk, which, for $\Lambda_B<0$, has the form
\be
a^2(t,y)=d_1(t)\,\cosh(A|y|) + d_2(t)\,\sinh(A|y|) - \frac{B^2(t)}{A^2}\,,
\label{general}
\ee
where
\be
A^2=\frac{2\hat \kappa^2}{3}\,|\Lambda_B|\,,\qquad
B^2(t)=\frac{2}{\lambda^2(t)}\,,
\ee
and where $d_1$ and $d_2$ are two functions of time which will
be shortly determined. The above solution, derived in the bulk, needs
to be smoothly connected to its boundary values $a_0$ and $a_L$. We can write
\be
{\rm \bf (A):}  \quad a^2(t,0)=a_0^2(t), \qquad a^2(t,L)=a_L^2(t)\,.
\ee
The first of these conditions leads to the determination of the unknown
function
$d_1(t)$,
\be
d_1(t)=a_0^2(t) + \frac{B^2(t)}{A^2}
\label{defd1}
\ee
while the second leads to the relation between the two boundary values
$a_0$ and $a_L$
\be
a_L^2(t)=d_1(t)\,\cosh(A L) + d_2(t)\,\sinh(A L) - \frac{B^2(t)}{A^2}\,.
\label{defaL}
\ee
Two more sets of boundary conditions need to be satisfied by the general
solution (\ref{general}). These conditions follow from the inhomogeneity
in the distribution of matter in the universe and involve the {\it jumps}
in the first derivatives of $a$ and $n$ across the two branes. They have
the form
\ba
&~& \hspace*{-0.5cm} {\rm \bf (B):}  \quad 
\frac{[a']_0}{a_0}=-\frac{\hat{\kappa}^2}{3}\,(\rho_1+ \Lambda_1)\,, 
\qquad 
\frac{[a']_L}{a_L}=-\frac{\hat{\kappa}^2}{3}\,(\rho_2+ \Lambda_2)\,,\\[2mm]
&~& \hspace*{-0.5cm} {\rm  \bf (C):}  \quad 
\frac{[n']_0}{n_0}=\frac{\hat{\kappa}^2}{3}\,
(3p_1+2\rho_1-\Lambda_1)\,, 
\qquad \frac{[n']_L}{n_L}=\frac{\hat{\kappa}^2}{3}\,
(3p_2+2\rho_2-\Lambda_2)\,.
\label{jumpsn}
\ea
We first focus on set ({\bf B}) : the condition at $y=0$ determines the
remaining arbitrary function $d_2(t)$ to be
\be
d_2(t)=-\frac{\hat{\kappa}^2}{3 A}\,a_0^2\,(\rho_1+\Lambda_1)\,,
\label{defd2}
\ee
while the same condition at $y=L$, combined with the relation (\ref{defaL}),
leads to two extremely interesting results: the Friedmann
equation at $y=0$,
\be
\biggl(\frac{\dot{a}_0}{a_0}\biggr)^2=\frac{\hat \kappa^2 |\Lambda_B|}{3}\,
\frac{\biggl[-\Lambda_{RS}+\frac{\textstyle (\rho_1+\Lambda_1+
\rho_2+\Lambda_2)}{\textstyle \tanh(AL)} -
\frac{\textstyle (\rho_1+\Lambda_1)(\rho_2+\Lambda_2)}
{\textstyle \Lambda_{RS}} \biggr]}
{\Lambda_{RS} - (\rho_2+\Lambda_2)\,\tanh(\frac{AL}{2})}\,,
\label{hubble0}
\ee
and the ratio of the scale factors on the two branes
\be
\frac{a_L^2}{a_0^2}=\frac{\Lambda_{RS} - (\rho_1+\Lambda_1)\,
\tanh(\frac{AL}{2})}
{\Lambda_{RS} - (\rho_2+\Lambda_2)\,\tanh(\frac{AL}{2})}\,,
\label{rela}
\ee
expressed in terms of $\Lambda_{RS} \equiv \sqrt{6 |\Lambda_B|/\hat \kappa^2}$,
$\Lambda_i$, $\rho_i$ and $L$. 
The above form of the Friedmann equation has also been
derived in Ref.~\cite{kim} while some of the ex\-pressions presented in this
section have been determined, in the special case
$\Lambda_1=-\Lambda_2=\Lambda_{RS}$, in Ref.~\cite{sissa}.
As in the case of static
solutions \cite{kop1}, the ratio of the scale factors on the two branes
is given in terms of the `detuning' of the total energy densities of
the branes from the limiting value $\Lambda_{RS}$. Differentiating the
above expression with respect to time, we are able to derive the
form of the Hubble parameter on the second brane, which is
\be
\biggl(\frac{\dot{a}_L}{a_L}\biggr)=
\biggl(\frac{\dot{a}_0}{a_0}\biggr)\,\Biggl[\frac{\Lambda_{RS}
- (\rho_2+\Lambda_2)\,\tanh(\frac{AL}{2})}
{\Lambda_{RS} - (\rho_1+\Lambda_1)\,\tanh(\frac{AL}{2})}\Biggr]
\Biggl[\frac{2\Lambda_{RS}
- (2\Lambda_1-\rho_1-3p_1)\,\tanh(\frac{AL}{2})}
{2\Lambda_{RS} - (2\Lambda_2-\rho_2-3p_2)\,\tanh(\frac{AL}{2})}\Biggr]\,.
\label{hubbleL}
\ee

Finally, let us note that the set ({\bf C}) of boundary conditions involving
the {\it jumps} of the lapse function $n$ across the two branes must also
be satisfied. By using the fact that $n(t,y)=\dot{a}(t,y)/\dot{a}_0(t)$,
differentiating with respect to $y$ and substituting in eqs. (\ref{jumpsn}),
we may see, after some algebra, that the jump conditions on both boundaries
are identically satisfied. The above result reveals the absence of any
correlation between the equations of state on the two brane-universes. This
feature was also pointed out in Ref. \cite{sissa}, in the special case
$\Lambda_1=-\Lambda_2=\Lambda_{RS}$, however, it turns out to be valid
even in the case where the self-energies of the branes and the bulk
cosmological constant are completely arbitrary and uncorrelated.

We now turn to the conditions that we need to impose in order to recover
the usual Friedmann equation on both branes. The most obvious is the
vanishing of the effective cosmological constant. From eq. (\ref{hubble0}),
this translates to 
\be
\Lambda_{eff} \sim \Lambda_{RS}\,\biggl( -1 - \alpha \beta + \frac{\alpha +
\beta}{\tanh(AL)}\biggr)=0\,,
\label{eff}
\ee
where we have defined
\be
\alpha \equiv \frac{\Lambda_1}{\Lambda_{RS}}\,, \qquad
\beta \equiv \frac{\Lambda_2}{\Lambda_{RS}}\,.
\ee
The condition (\ref{eff}) involves three parameters, $\alpha$, $\beta$ and
$L$ which, until now, were completely uncorrelated. The condition of the
vanishing of $\Lambda_{eff}$ renders one of them a dependent parameter,
thus, reducing the number of independent parameters to two. Note that
there is a special choice for the parameters $\alpha$ and $\beta$ that
eliminates $\Lambda_{eff}$, for every value of the inter-brane distance
$L$. This is $\alpha=-\beta=\pm 1$ and corresponds to the Randall-Sundrum
choice~\cite{RS1}. For every other combination of $\alpha$ and $\beta$,
the distance between the two branes has to be carefully chosen in order
to ensure that $\Lambda_{eff}=0$. 

Although the vanishing of the effective cosmological constant has
simplified the Fried\-mann-like equation at $y=0$,
eq.~(\ref{hubble0}), it has not completely restored its usual form. 
The existence of two branes with non-vanishing energy densities,
$\rho_i$, leads to the appearance of both $\rho_1$ and $\rho_2$, in the
equations that govern the cosmological evolution of each brane. 
Besides terms linear in  $\rho_1$ and $\rho_2$, the resulting cosmological 
equations will be corrected by higher powers of energy densities, i.e.
$\rho_1^2$, $\rho_2^2$, $\rho_1\rho_2$, and so on. 
Exact numerical coefficients in front of these terms are 
not known, as they depend on subtle 
details of radius stabilization \cite{kkop2}. However, for a natural choice
of parameters, 
these terms can easily be shown
to be of secondary importance due to their extremely small magnitude.
If we assume that $\Lambda_{RS} \sim M_P^4$ while $\rho_i \la \rho_c
\simeq 10^{-123} M_P^4$ (as dictated by the observable expansion rate
of our 4D universe), and expand eq. (\ref{hubble0}) keeping only terms
linear in $\rho_i/\Lambda_{RS}$, we obtain the result
\be
\biggl(\frac{\dot{a}_0}{a_0}\biggr)^2=\frac{\hat \kappa^2 |\Lambda_B|}
{3 \Lambda_{RS} \tanh(AL)}\,\Biggl\{
\frac{1-\beta \tanh(AL)}{1-\beta \tanh(\frac{AL}{2})}\,\rho_1 +
\frac{1-\alpha \tanh(AL)}{1-\beta \tanh(\frac{AL}{2})}\,\rho_2
\Biggr\}\,.
\label{hubble02}
\ee
We could also use the condition of the vanishing of $\Lambda_{eff}$
in order to eliminate one of the three parameters $\alpha$, $\beta$
and $L$. We have chosen to eliminate $\alpha$, using (\ref{eff}).
The final
form of the Friedmann equation, at $y=0$, then takes the form
\be
\biggl(\frac{\dot{a}_0}{a_0}\biggr)^2=\frac{\hat \kappa^2 |\Lambda_B|}
{3 \Lambda_{RS} \tanh(AL)}\,\frac{1-\beta \tanh(AL)}
{1-\beta \tanh(\frac{AL}{2})}\Biggl\{\rho_1 +
\frac{\rho_2}{[\cosh(AL)-\beta \sinh(AL)]^2}\Biggr\}\,.
\label{hubble03}
\ee
Before commenting on the above result, we should point out that the
Friedmann equation on the observable brane, at $y=L$, has exactly
the above form, in the same linear approximation. We can easily see,
from eq. (\ref{hubbleL}), that by ignoring terms of ${\cal O}
(\rho_i/\Lambda_{RS})^2$ or ${\cal O}(\rho_i \rho_j/\Lambda_{RS}^2)$,
the two Hubble parameters become identical. In
eq.~(\ref{hubble03}) we would restore the normal 4-dimensional Hubble
expansion law if we define Newton's constant as
\be
\kappa^2= \frac{\hat \kappa^2\,|\Lambda_B|}
{\Lambda_{RS} \tanh(AL)}\,\frac{1-\beta \tanh(AL)}
{1-\beta \tanh(\frac{AL}{2})}\,.
\ee
The square of the Hubble parameter is now proportional to the
total energy density of the universe, which contains contributions
from both branes. Note that the constant
coefficient in front of the energy density of our brane-universe,
$\rho_2$, is not unity (this was first pointed out in Ref. \cite{randall3}).
As
was discussed in~\cite{sissa,mpp}, the observable energy density on our
universe must be redefined in such a way as to absorb this coefficient.
In the next section we will see that this redefinition is a result of the 
conformal transformation necessary to solve the hierarchy problem.

Finally, we need to address the problem of the stabilization of the
extra dimension and determine the constraints on the energy distribution
of the system that makes this stabilization possible. As we mentioned before
and demonstrated in earlier works~\cite{kkop1,kkop2}, the stabilization
mechanism responsible for keeping fixed the inter-brane distance manifests
itself through the existence of a non-vanishing (55)-component of the
energy-momentum tensor. The (55)-component of Einstein's equations,
that we have ignored so far, will serve to determine the expression
of this extra component. For the general solution (\ref{general}), 
eq. (\ref{55}) takes the form
\be
\frac{d}{dt}\,\Bigl[\frac{A^2}{4}\,(d_2^2-d_1^2) +
\frac{1}{A^2 \lambda^4}\Bigr]=
\frac{2\hat \kappa^2}{3}\,a^3 \dot{a}\,(\hat T^5_{(B)\,5}-|\Lambda_B|)\,.
\ee
Substituting $d_1$, $d_2$ and $\lambda$ from eqs. (\ref{defd1}),
(\ref{defd2}) and (\ref{hubble0}), respectively, we obtain the result
\ba
\hat T^5_{(B)\,5} &=& |\Lambda_B| + \frac{a_0^3\,\hat\kappa^2}{12 n a^3}
\,\Biggl\{(\rho_1+\Lambda_1)\,(2\Lambda_1-\rho_1-3p_1) +\Lambda_{RS}\,
\Bigl[2 \Lambda_{RS}-\frac{(4\Lambda_1+\rho_1-3p_1)}{\tanh(AL)}\Bigr]
\nonumber \\[1mm]
&-&\frac{\Lambda_{RS}}{\sinh(AL)}\,\frac{a_L}{a_0}\,\Bigl[
\,2\frac{a_L}{a_0}\,(\rho_2+\Lambda_2) + n_L\,(2\Lambda_2 -3 p_2
-\rho_2)\,\Bigr]\Biggr\}\,,
\label{res55a}
\ea
and to linear order in $\rho_i$ 
\ba
\hat T^5_{(B)\,5} &=& |\Lambda_B| + \frac{a_0^3\,|\Lambda_B|}{2 n a^3}
\,\Biggl\{\frac{2\,(\beta^2-1)}{\cosh^2(AL)\,[1-\beta\tanh(AL)]^2}
\nonumber \\[2mm]
&-&\frac{(\rho_1-3p_1)\,[1-\beta\tanh(\frac{AL}{2})] +  (\rho_2-3p_2)\,
[1+\beta\tanh(\frac{AL}{2})]}{\Lambda_{RS}\,\sinh(AL)\,\cosh(AL)\,
[1-\beta\tanh(AL)]}\Biggr\}\,.
\label{res55b}
\ea
Note that this expression already assumes the vanishing of the
cosmological constant (\ref{eff}). We may easily check that the expression
(\ref{res55a}), as well as the linearized version above, with $\hat
T^5_{(B)\,5}
\sim w(t)/n(t,y)\,a^3(t,y)$, where $w(t)$
an arbitrary function of time, is consistent with the fifth component of the
equation for the conservation of energy (\ref{fifth}). In addition, as we
expected, the coefficient $w$ is proportional to the trace of the energy
momentum tensor on the branes \cite{kkop2,ell}. Note also that we neglect
any small changes in $b$, due to changes in $\hat T^5_{(B)\,5}$ or $\rho_i$,
since these are proportional to $\rho^{3/2}_i$ and we consistently ignore
terms of this order throughout our analysis.


\sect{Resolution of the Hierarchy Problem}

After having stabilized the extra dimension and derived the Friedmann
equations that govern the cosmological evolution of each brane, we now
focus our attention on the observable brane-universe. Our goal will be
to determine the correct definition of mass scales, $\tilde m$, and
energy density, $\tilde \rho_2$, as measured by a 4D observer in a FRW
background, in terms of the corresponding quantities, $m$ and
$\rho_2$, defined on our 4-brane which, however, is part of a curved,
non-FRW spacetime. (In this context, by FRW we refer to a metric with lapse
function $n=1$, with only a time dependent spatial scale factor.)  To this
end, we consider the theory of a scalar field,
$\Phi$, confined on our brane-universe which is embedded in a 5D spacetime
background. Such a toy model was also considered in Ref. \cite{RS1}, however,
here, we extend it to the case of a non-static spacetime background of the
form (\ref{metric}). We consider the following action
\ba
S&=& \int d^4x\,dy\,\sqrt{\hat g}\,\biggl[\,\hat g^{\mu\nu}\, 
\partial_\mu \Phi\,\partial_\nu \Phi + \lambda\,(\Phi^2-v_0^2)^2\biggr]
\delta(y-L) \nonumber \\[2mm]
&=&  \int d^4x\,\sqrt{g_L}\,\biggl[\,g_L^{\mu\nu} 
\partial_\mu \Phi\,\partial_\nu \Phi + \lambda\,(\Phi^2-v_0^2)^2\biggr]\,,
\label{5D}
\ea
where $g_{L\mu\nu}=\hat g_{\mu\nu}(y=L)$, is the induced metric tensor
on our brane. Using the ansatz (\ref{metric}), the above action
takes the form
\ba
S &=& \int d^4x\, (a_L^3 n_L)\,\biggl\{ \Bigl[-\frac{1}{n_L^2}\,
\dot \Phi^2+ \frac{1}{a_L^2}\,(\nabla \Phi)^2\Bigl] 
+ \lambda\,(\Phi^2-v_0^2)^2\biggr\} \nonumber \\[2mm]
&=& \int d^4x\,\sqrt{\tilde g}\,n_L^4\,\biggl[ \frac{1}{n_L^2}\,
\tilde g^{\mu\nu} \partial_\mu \Phi\,\partial_\nu \Phi +
\lambda\,(\Phi^2-v_0^2)^2\biggr]\,,
\ea
where a conformal transformation
\be
g_{L\mu\nu} \rightarrow \tilde g_{\mu\nu}= \frac{1}{n_L^2}\,g_{L\mu\nu}
\label{conformal}
\ee
has been performed that restores the FRW character of the 4D
spacetime and brings the corresponding line-element to the form
\be
d\tilde s^2=-dt^2 + \tilde a^2(t)\,\delta_{ij} dx^i dx^j\,,
\ee
where $\tilde a(t)\equiv a_L(t)/n_L(t)$ is the observed scale
factor of our universe. The last step we need to make 
involves the rescaling of the scalar field in order to absorb the
gravitational coefficients that remain in the action. By setting
$\phi \rightarrow \tilde \phi = n_L\,\phi$, we arrive at 
\be
S = \int d^4x\,\sqrt{\tilde g}\,\bigl[\,
\tilde g^{\mu\nu} \partial_\mu \tilde \Phi\,\partial_\nu \tilde \Phi +
\lambda\,(\tilde\Phi^2-\tilde v_0^2)^2\bigr]\,.
\label{4D}
\ee
The $vev$ $v_0$ of the scalar field has also been rescaled, $\tilde v_0 = n_L
v_0$, and as a result, the relation between any mass scale,
$\tilde m$, defined in terms of the theory (\ref{4D}) and the corresponding
one, $m$, defined in terms of (\ref{5D}), has the form
\be
\tilde m = n_L(t)\,m \,.
\label{masses}
\ee
We may, thus, conclude that the rescaling coefficient that relates any
effective mass scale with the corresponding fundamental one, in a 
general spacetime background, is the lapse function of the original
5D line-element evaluated at the location of the brane. In the case
of the RS model, $n_L=e^{-bkL}$ and thus any mass scale $\tilde m$ is
exponentially suppressed compared to $m$. Due to the fact that $n_L=a_L$,
the induced line-element, in their case, was static since
$\tilde a(t) = a_L/n_L= 1$. In our case, however, the presence of
the energy density $\rho_2$ inevitably leads to the expansion of our 
brane and to the time-dependence of the rescaling factor $n_L$ . In Ref. 
\cite{sissa}, it was argued that such a rescaling is not acceptable
since the time-dependence of $n_L$ would modify the equations of 
motion of the scalar field $\Phi$. Since $\dot{\tilde \Phi} = \dot n_L
\Phi$, the above problem is resolved only if
\be
\frac{\dot{\tilde \Phi}}{\tilde \Phi}= \frac{\dot n_L}{n_L} \ll 1\,.
\label{timecon}
\ee
Moreover, any significant time evolution of the conformal factor $n_L$
would lead to the time-dependence of either the fundamental mass scale,
$m$, or the effective one, $\tilde m$.

It is straightforward to check that the above condition is indeed satisfied
in our case. The expression for the lapse function at $y=L$ can be obtained
using the relation $n_L(t)=\lambda(t)\,\dot a_L(t)$ and eq.~(\ref{hubbleL}),
and is found to be
\be
n_L(t)=\frac{a_0}{a_L}\,\,
\frac{2\Lambda_{RS}-(2\Lambda_1 -3p_1-\rho_1)\,\tanh(\frac{AL}{2})}
{2\Lambda_{RS}-(2\Lambda_2-3p_2 -\rho_2)\,\tanh(\frac{AL}{2})}\,.
\label{defnL}
\ee
Differentiating the above expression with respect to time, we find
\be
\frac{\dot n_L}{n_L}= \frac{\dot a_0}{a_0} - \frac{\dot a_L}{a_L}
+ \frac{(3w_1+1)\,\dot \rho_1\,\tanh(\frac{AL}{2})}
{2 \Lambda_{RS}\,[1-\alpha \tanh(\frac{AL}{2})]}-
\frac{(3w_2+1)\,\dot \rho_2\,\tanh(\frac{AL}{2})}
{2 \Lambda_{RS}\,[1-\beta \tanh(\frac{AL}{2})]}\,,
\ee
where we have set $p_i=w_i\,\rho_i$ and ignored higher-order terms
of order ${\cal O}(\rho_i \dot \rho_i)$. As we have shown in the previous
section, the first two terms are identical if we ignore corrections
of order ${\cal O}(\rho_i/\Lambda_{RS})^2$, which are extremely small
if we recall that $\rho_i/\Lambda_{RS}\sim 10^{-123}$. The remaining
two terms can be easily shown to be proportional to
$\dot \rho_i/\Lambda_{RS} \sim (\rho_i/\Lambda_{RS})^{3/2} M_P$
which undoubtly proves the validity of the condition (\ref{timecon}).
It may also be worth checking the validity of this argument
at the time of the Big-Bang Nucleosynthesis. The corresponding term
in the expression of $\dot n_L$ would be $10^{-148}
M_P\,(\rho_N/\rho_c)^{3/2}$, where $\rho_N$ is the energy density of the
universe during the BBN.  If we assume that $T_{BBN} \simeq 10$ MeV, we
may easily find that 
$\rho_N/\rho_c \simeq 10^{39}$. This, in turn, leads to the final
result that the magnitude of the term $\dot \rho_i/\Lambda_{RS}$
at the same period was of the order of $10^{-90} M_P$ rendering the
time-dependence of the lapse function $n_L$ completely unobservable
even in the early universe.

  If we assume
that the fundamental energy scale of the 5D theory in eq. (\ref{masses}), $m$,
is of order
$M_P$, while $\tilde m \sim 1$ TeV,
the rescaling coefficient should be of order $\sim 10^{-16}$. 
Expanding the result  for the lapse function (\ref{defnL}) and keeping
only terms linear in
$\rho_i/\Lambda_{RS}$, we obtain
\be
n_L=\sqrt{\frac{1-\alpha \tanh(\frac{AL}{2})}{1-\beta \tanh(\frac{AL}{2})}}
\, \Biggl\{1+ \frac{(3w_1+2)\,\rho_1\,\tanh(\frac{AL}{2})}
{2 \Lambda_{RS}\,[1-\alpha \tanh(\frac{AL}{2})]}-
\frac{(3w_2+2)\,\rho_2\,\tanh(\frac{AL}{2})}
{2 \Lambda_{RS}\,[1-\beta \tanh(\frac{AL}{2})]}\Biggr\}\,.
\label{expnL}
\ee
Since, the last two terms inside the
brackets are negligible compared to unity, the condition that 
$n_L \simeq 10^{-16}$ should be imposed on the prefactor appearing in
(\ref{expnL}) involving $\alpha$, $\beta$ and $L$.
Note that this prefactor is nothing more than the ratio of the
scale factors on the two branes (\ref{rela}), once the small energy density
contributions are ignored. By making use of eq. (\ref{eff}),
this condition can be conveniently rewritten as
\be
n_L^2 =\frac{1}{\cosh(AL) - \beta\,\sinh(AL)} \simeq 10^{-32}\,,
\label{conv1}
\ee
Once eq. (\ref{conv1}) is imposed, we are left with only one independent
parameter in the theory which we choose to be $L$, the inter-brane distance.
The parameter $\beta$ can be determined by the condition
(\ref{conv1}), which together with the tuning of the cosmological
constant (\ref{eff}) give the following expressions for $\alpha$ and
$\beta$ in terms of $L$
\be
\alpha = \frac{\cosh(AL) - 10^{-32}}{\sinh(AL)}\,,
\label{eff1}
\ee
and 
\be
\beta=\frac{\cosh(AL) - 10^{32}}{\sinh(AL)}\,.
\label{conv2}
\ee
According to our notation, 
\be
A L= \sqrt{\frac{2}{3}\,\frac{|\Lambda_B|}{M_5^3}}\,L 
\simeq M_P\,L\,,
\ee
if we assume that $|\Lambda_B|^{1/5} \simeq M_5 \simeq  M_P$. In that case, 
the quantity $AL$ is nothing more than the distance between
the two branes measured in units $M_P^{-1}$. Since the inter-brane
distance is the free parameter of the theory, it would be interesting
to see how the dependent parameters $\alpha$ and $\beta$ change as
we vary $A L$. Their values will be dictated by eqs. (\ref{eff1})
and (\ref{conv2}), respectively, if we demand that both of the
cosmological and the hierarchy problems are simultaneously solved.
Numerical results for $\alpha$ and $\beta$ are plotted in Figure 1 as a
function of $AL$ and some specific numerical examples are given in
Table I.

\begin{center}
{\small {\bf Table I~:} The parameters $\beta$ and $1-\alpha$
for various values of the inter-brane distance $L$.}\\[2mm]
$
\begin{array}{|c|c|c|c|c|} \hline \hline
{\bf A L} & {\bf 1} &  {\bf 20} &  {\bf 40}  & {\bf 60} \\ \hline
\beta & -8.5092 \times 10^{31} & -9.0799 \times 10^{27} & 
-8.4967 \times 10^{14} & -1.7513 \times 10^6 \\ \hline
1-\alpha & -0.31304 & -4.1223 \times 10^{-9}& -3.6097 \times 10^{-35} 
& -1.5335 \times 10^{-52}  \\ \hline \hline
{\bf A L}  & {\bf 70} & {\bf 75} & {\bf 80} & {\bf 85}\\ \hline
\beta &  -78.509 & 0.46427&  0.99639 & 0.99998\\ \hline
1-\alpha & -3.0814 \times 10^{-61} & 3.9223 \times 10^{-65}
& 3.6032 \times 10^{-67} & 1.6388 \times 10^{-71} \\ \hline \hline
\end{array}$
\end{center}

\begin{figure}
\begin{center}
\mbox{\epsfig{file=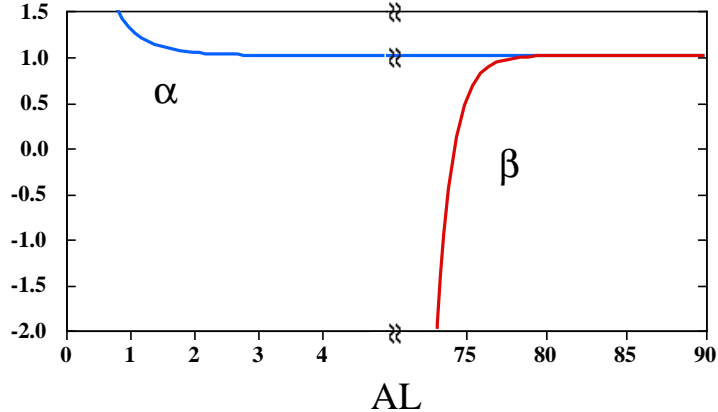,height=6cm}}
\end{center}
\caption[.]{\label{fig:bsg1}\it
Behavior of the tensions (in units of $\Lambda_{RS}$), $\alpha$ and
$\beta$ as a function of the inter-brane separation (in units of
$A^{-1}$).}
\end{figure}

We note that, for small values of the free parameter $L$, the brane
parameter $\alpha$ is positive and of ${\cal O}(1)$ while $\beta$ is
forced to acquire a value which is, not only negative, but also many
orders of magnitude larger than that of $\alpha$. Such a situation is
rather unphysical as it leads to an observable brane-universe with
negative, total energy density and to an unexplained hierarchy between
the sizes of the brane tensions. As $L$ increases, however, both 
$\alpha$ and $\beta$ start approaching unity: the combination
$1-\alpha$ becomes extremely small while $\beta$ is moving towards
less negative values. At $L \simeq 73.682722...$, the critical point
$\beta=-1$ and $\alpha \simeq 1$, the choice of brane parameters in
the Randall-Sundrum\footnote{Note, that our conventions differ from
those in Ref. \cite{RS1} by a factor of 2. Thus, their warp factor
$e^{-kr_c\pi}$ corresponds to $e^{-AL/2}$, in our case, if we ignore
the matter on the brane and set $\alpha=-\beta=1$.} model~\cite{RS1},
is reached. If we further increase $L$, the combination $1-\alpha$
changes sign but it keeps approaching zero. On the other hand, $\beta$
becomes positive at a slightly
larger value of the inter-brane distance, at $A L \simeq 75$. Any
further increase in the value of $AL$ forces both of the brane
parameters, $\alpha$ and $\beta$, to unity, which is the largest
value that these parameters may take on. This is clear from eq. (\ref{eff}), 
for $\alpha=\beta=1$, where the hyperbolic tangent takes its asymptotic
value, i.e. $\tanh(AL)=1$, and corresponds to an infinite distance between
the two branes. Once the two branes become isolated, the only solution
to the cosmological constant problem follows for $\alpha=\beta=1$.
This is obvious from eq. (\ref{hubble0}), which in the limit
$\tanh(AL) \rightarrow 1$, takes the form
\be
\biggl(\frac{\dot{a}_0}{a_0}\biggr)^2=
\frac{\hat \kappa^2 |\Lambda_B|}{3}\,
\biggl[-1+\frac{\textstyle (\rho_1+\Lambda_1)}
{\textstyle \Lambda_{RS}}\biggr]\,.
\ee
In this limit, the expansion of the brane at $y=0$ becomes independent
of the presence of the second brane at $y=L$ and the flatness of the
corresponding 4D spacetime is guaranteed only for $\alpha=1$ (similar
results hold for the observable universe at $y=L$).
Thus, for any value of the inter-brane distance,
larger than, approximately, $75\,M_P^{-1}$, the brane parameters
are always positive, with their upper limit being unity,
a result which leads to two brane-universes
with positive, total energy densities. Clearly, if $A L \gg 75$,
then, in addition to tuning $1-\alpha$ to zero, $\beta$ must also be
tuned to unity. We should stress here, once
again, that all of the above values of $\alpha$ and $\beta$, either
positive or negative, are
in accordance with both the vanishing of the effective cosmological
constant on both branes and with the resolution of the hierarchy
problem on the observable universe.

Let us finally address the problem of the definition of the energy
density, $\tilde \rho_2$, on our brane as measured by an observer 
living in a FRW spacetime in terms of the original
energy density, $\rho_2$, defined in a 4D sub-space with lapse function
$n_L \ne 1$. The energy-momentum tensor associated with the 
matter content of our universe is defined as 
\be
T^M_{\mu\nu}=\frac{2}{\sqrt{g_L}}\,\frac{\delta S^M}{\delta
g_L^{\mu\nu}}\,,
\ee
where $S_M$ is the part of the original action (\ref{action0}) defined
only on our brane and containing time-dependent functions that describe
the distribution of matter sources in our universe. After the conformal
transformation (\ref{conformal}), $T^M_{\mu\nu}$ changes as~\cite{wetterich}
\be
T^M_{\mu\nu} \rightarrow \tilde T^M_{\mu\nu}=\frac{2}{\sqrt{\tilde g}}\,
\frac{\delta S^M}{\delta \tilde g^{\mu\nu}}= n_L^2\,
T^M_{\mu\nu}\,.
\ee
By using the fact that $\tilde T^\mu_\nu=(-\tilde \rho_2, \tilde p_2,
\tilde p_2, \tilde p_2)$ and $\tilde T^\mu_\nu=\tilde g^{\mu\rho}
\tilde T_{\rho\nu}=n_L^4 T^\mu_\nu$, we finally obtain the results
\be
\tilde \rho_2 = n_L^4\, \rho_2\,, \qquad \tilde p_2 = n_L^4\, p_2\,.
\label{redef}
\ee
The above definition of the observed energy density $\tilde \rho_2$
is in perfect agreement with the form of the Friedmann equation
(\ref{hubble03}) for the cosmological expansion of our brane-universe.
Recall that, to linear order in $\rho_i$, the expansion rate is 
\be
\biggl(\frac{\dot{a}_0}{a_0}\biggr)^2 \propto (\rho_1 + n_L^4 \rho_2) =
(\rho_1 + \tilde \rho_2)\,,
\label{finalh}
\ee
when one uses the expression for $n_L$ (\ref{conv1}) in  terms of
the brane parameter $\beta$ and the inter-brane distance $AL$ (here,
we consider only the lowest-order expression of $n_L$ since any
$\rho_i$-dependent terms would create higher-order corrections,
of ${\cal O}(\rho_i^2, \rho_i\rho_j)$, in the Friedmann equation).
Note that no redefinition is necessary for the components of the
4D energy-momentum tensor on the brane at $y=0$ since the lapse
function, and thus the conformal factor, $n_0$, is unity. Similarly, 
we have $\tilde a_0 =a_0$, and, as a result,  eq. (\ref{finalh})
still governs the expansion of the 4D FRW spacetime at $y=0$. Although
the rescaling of the scale factor  at $y=L$ is not trivial, we can easily
see that
\be
\frac{\dot {\tilde a}_L}{\tilde a_L} =
\frac{ \dot a_L}{a_L}-\frac{\dot n_L}{n_L}=
\frac{ \dot a_0}{a_0} + {\cal O}(\frac{\rho_i^2}{\Lambda^2_{RS}},
\frac{\rho_i\rho_j}{\Lambda^2_{RS}})\,.
\ee
Therefore, in the linear-order approximation in the energy densities,
the same equation (\ref{hubble03}) describes the expansion of the
4D FRW spacetime at $y=L$. After performing the redefinition
(\ref{redef}), the unique Friedmann equation (\ref{finalh}) reveals
the fact that the cosmological evolution of both branes is governed by
the sum of the energy densities of the two branes as measured by observers
living in flat, FRW 4D spacetimes. The same conclusion regarding
the Friedmann expansion of both branes was also derived in Ref.
\cite{sissa} in the special case of the RS choice of the brane parameters
$\alpha$ and $\beta$. 

\section{Conclusions}

Before summarizing the work in this paper, let us emphasize briefly the main
features of the two-brane models which may naturally explain the large
ratio $\sim M_{P}/M_W$ via the hierarchy of the two length scales on
different branes.  In the RS model \cite{RS1}, the solution for the
warp factor in the bulk, $a(y)$, was given by an exponential function
of $y$ that interpolated between a positive and a negative self-energy
brane. In the two-positive-brane model \cite{kop1}, the corresponding
solution behaved like a $\cosh$-function characterized by the existence
of a minimum. The solution to the hierarchy problem may be explained
by the position of these two positive branes with respect to the minimum.
Indeed, as we have shown, the ratio of the two lapse functions can be large,
when the branes are located at non-equal distances from the minimum. 

In this work, we have generalized these static models by introducing energy
densities on the branes. Moreover, we have allowed for the possibility of
arbitrary (positive or negative) brane tensions. We have
shown that models of this type exhibit ``normal'' cosmology after demanding
the vanishing of the effective cosmological constant on both branes and
redefining the energy density on the observable brane in order to restore
the FRW character of our 4D universe. It is also worth noting that no
correlation between states of matter on the respective branes is required. 
In addition, we have shown that despite the time-dependence of the lapse
function on the observable brane, that defines the ratio $M_{P}/M_W$,
its time evolution does not lead to any measurable change of the effective
four-dimensional parameters since such changes would be proportional to
$\rho_i/M_P^4$.

A crucial element in our model is the mechanism, responsible for the 
stabilization of the inter-brane distance. Here we simply assume that this
mechanism exists without specifying physical causes, leading to the 
radion/dilaton stabilization. An
extended discussion of this issue can be found in  Ref.
\cite{kkop2}.  Of course, any model with a string like
dilaton must ensure such a stabilization, since the dilaton expectation
value fixes the gauge coupling and mass scales in the standard model.  

We can not claim that two-brane models of this type are devoid of 
fine-tunings.  As in all other known models, the cosmological
constant requires a fine-tuning of different parameters in
the model to ensure $\Lambda_{eff} = 0$. This reduces the initial set
of three parameters (the two brane tensions relative to the bulk
cosmological constant, and the inter-brane separation) to two. Note
that the bulk cosmological constant relates the 4D and 5D (fundamental)
Planck scales. When $\Lambda_{eff}$ is set to zero, the weak scale/Plank
scale hierarchy problem may be explained by fixing a relation between the
remaining brane tension and the inter-brane separation. For every value
of the inter-brane distance, a solution to both the cosmological constant
problems and the hierarchy problem can be derived by appropriately
choosing the tension of the observable universe.  While for small
values of $AL$, the brane parameter $\beta$ is forced to acquire a large
negative value, as the inter-brane distance increases, both of the
parameters $\alpha$ and $\beta$ start decreasing (in absolute value)
towards unity. The RS choice of parameters, i.e. $\alpha=-\beta=1$, arises
as a special solution when we reach the value $AL \simeq 73$ while, for
all separations
$AL \ga 75$, more natural solutions (with positive brane tensions for
both branes) arise. The parameters $\alpha$ and $\beta$ remain positive
no matter how large the inter-brane separation becomes. The asymptotic
value $\alpha=\beta=1$ is reached in the limit of infinite $AL$, when
the two branes become isolated. In this limit, the problem of the
cosmological constant is solved by imposing independent conditions on
each brane and the framework for the resolution of the hierarchy
problem ceases to exist.

{\bf Acknowledgments} This work was supported in part by
the Department of Energy
under Grant No.\ DE-FG-02-94-ER-40823 at the University of Minnesota.


\end{document}